\documentstyle[twocolumn,aps,psfig]{revtex}
\begin{document}
\draft

%
\twocolumn[\hsize\textwidth\columnwidth\hsize\csname@twocolumnfalse\endcsname
%
%

\title{Static and Dynamical Properties of the Ferromagnetic Kondo Model with
Direct Antiferromagnetic Coupling Between the localized $t_{2g}$ Electrons}

\author{ Seiji Yunoki and Adriana Moreo}

\address{Department of Physics, National High Magnetic Field Lab and
MARTECH, Florida State University, Tallahassee, FL 32306, USA}

\date{\today}
\maketitle

\begin{abstract}
The phase diagram of the Kondo lattice Hamiltonian with ferromagnetic
Hund's coupling in the limit where the spin of the localized
$t_{2g}$ electrons is classical is analyzed in one dimension as a function of
temperature, electronic density, and a direct
antiferromagnetic coupling $J'$ between 
the localized spins. Studying static and dynamical properties, a behavior
that qualitatively resembles experimental results for 
manganites occurs for $J'$ smaller than 0.11 in units of the $e_g$
hopping amplitude. In particular a coexistence of ferromagnetic and
antiferromagnetic excitations is observed at low-hole density
in agreement with neutron
scattering experiments on $\rm{La_{2-2x}Sr_{1+2x}Mn_2O_7}$ with
$x=0.4$. This effect is caused by the recently 
reported tendency to phase separation between hole-rich ferromagnetic
and hole-undoped antiferromagnetic domains in electronic models for
manganites.
As $J'$ increases metal-insulator transitions are detected by monitoring
the optical conductivity and the density of states. The magnetic
correlations reveal the existence of spiral phases without long-range
order but with fairly large correlation lengths. Indications of 
charge ordering effects appear in the analysis of charge correlations.
\end{abstract}

\vskip2pc]
\narrowtext

\section{Introduction}

The study of manganese oxides with perovskite-type structure such as
$\rm{La_{1-x}Ca_xMnO_3}$ and $\rm{La_{1-x}Sr_xMnO_3}$ has received new emphasis
since the discovery of colossal magnetoresistance in these
compounds~\cite{jin}. Experiments have revealed a rich phase diagram
with antiferromagnetic insulating, ferromagnetic
metallic, and charge ordered 
regions~\cite{schi,chen}. 
Recently, layered compounds with
properties similar to the 3D materials have been synthesized: the new compounds
are described by the general formula
$\rm{(A,B)_{n+1}Mn_nO_{3n+1}}$, where $n$ is the number of $\rm{Mn-O}$ planes per
unit cell ($n=\infty$ corresponds to the 3D case)~\cite{moh}. 
The appearance of ferromagnetism upon hole doping away from
$x=0$ has been attributed to the
double-exchange (DE) mechanism between the $e_g$ and $t_{2g}$ 
electrons~\cite{zen,ander,gen}. However, this model is
not sufficient to describe the rich phase diagram observed
experimentally. 
The large regions in the phase diagram which present
charge ordering are not contained in the DE model.
Features such as the antiferromagnetic phase in the
low $e_g$-electronic density
limit ($x \sim 1$), or the related experimentally observed
antiferromagnetic ground state detected in $\rm{La_{1-x}Ca_xMnO_3}$ for
$x>0.55$ are difficult to explain by DE ideas.
The N\'eel temperature of this state
reaches a maximum value of 260K and
decreases to 120K at $x=1$~\cite{schi}.

The main goal of this paper will be to study the phase diagram of a  modified
ferromagnetic (FM)
Kondo lattice Hamiltonian in order to understand qualitatively
what properties of the
manganites are solely due to electronic interactions. An analysis
including Jahn-Teller
phonons~\cite{mil} will be postponed for a future publication.
To achieve our goal 
the ferromagnetic Kondo lattice 
Hamiltonian will be here modified by introducing
a direct antiferromagnetic exchange interaction, $J'$, between the
spins of the localized $t_{2g}$ electrons~\cite{inoue,yuno}. 
This coupling will certainly produce an antiferromagnetic phase at
$x=1$, as experimentally observed,
which does not appear in the standard
FM Kondo model. The addition of
this coupling is certainly not unphysical since it
may originate from small hopping amplitudes for the $t_{2g}$
electrons. Experimentally such an effective antiferromagnetic exchange
interaction $J'$ has been studied in the 2D compounds and $2J'S\approx
100K$ has been estimated~\cite{aep}. 

Among the main issues that will be 
addressed in the present manuscript
is the influence of the direct coupling $J'$ among the $t_{2g}$ electrons
on the recently reported regime of {\it phase separation} detected in the
ferromagnetic Kondo model~\cite{yuno}.
This effect, as it will be discussed later, could be a key
feature to understand some of the recent neutron scattering results for the
manganites.

The organization of the paper is the following: in Section II the Hamiltonian
will be introduced. Phase separation will be studied in Section III.
Magnetic and charge ordering will be discussed in Section IV.
In Section V the existence of a Fermi surface is studied by monitoring
$n(k)$ and the spectral function $A(k,\omega)$. The optical conductivity,
the Drude weight and metal-insulator transitions are presented in
Section VI. In Section VII the phase diagram is proposed.
Section VIII is devoted to summarizing the results.

\section {The Model}

The parent
compound $\rm{LaMnO_3}$ contains $\rm{Mn^{3+}}$ ions in a $t^3_{2g}e^1_g$
configuration. The $t^3_{2g}$ electrons may be viewed as a localized
$S=3/2$ spin, while the $e^1_g$ electrons are mobile. Here one single orbital
for the $e_g$ electrons will be considered.
To describe the doped manganites a Kondo lattice model with
ferromagnetic Hund coupling has been proposed~\cite{kubo,furu}. In spite of
its apparent simplicity, this model is still difficult to study
accurately. In order to
simplify its analysis the localized $S=3/2$ spins will be replaced by classical
spins with $|{\bf{S}}_i|=1$~\cite{foot}. In addition, an antiferromagnetic 
coupling $J'$ will be
added between the classical spins to obtain the
antiferromagnetic phase at $x=1$ as explained in the 
Introduction~\cite{inoue,yuno}.
Finally, since the standard FM Kondo Hamiltonian in 3D is difficult to study
numerically due to its large number of degrees of freedom,
some other simplifications must be introduced.
Since previous studies have convincingly shown that the basic
features of the phase diagram
in three dimensional problems appear in two and one 
dimensions as well~\cite{yuno},
 results in one dimension will be here presented. 
A careful study of the 1D problem in chains with up to 40 sites will be
carried out using Monte Carlo (MC) techniques.

Then, the Hamiltonian studied in this paper is given by
$$
{ H=}
-t{ \sum_{\langle ij \rangle,\alpha}(c^{\dagger}_{i,\alpha}
c_{j,\alpha}+h.c.)}
$$

$$
-J_H
\sum_{i,\alpha,\beta}c^{\dagger}_{i,\alpha}{\bf{\sigma}}_{\alpha,\beta}c_{i,\beta}\cdot
{\bf{S}}_i+J'\sum_{\langle ij \rangle}{\bf{S}}_i \cdot{\bf{S}}_j,
\eqno(1)
$$

\noindent where ${ c^{\dagger}_{i,\alpha} }$ creates an $e_g$ electron
at site $i$ with spin projection $\alpha$, ${\bf{\sigma}}$ are Pauli
matrices, ${\bf{S}}_i$ is the
total spin of the $t_{2g}$ electrons assumed localized and classical,
 the sum ${ \langle ij \rangle }$ runs over pairs of nearest-neighbor 
lattice sites,
$t$ is the nearest-neighbor hopping amplitude for the $e_g$ electrons,
$J_H>0$ is the Hund coupling, and $J'>0$ is an antiferromagnetic coupling
between the localized spins. The density $\langle n \rangle=1-x$ of 
itinerant $e_g$-electrons is
controlled by a chemical potential $\mu$. 
In the following $t=1$ will be used as the unit of energy
and $J_H$ will be fixed at 8 since phenomenologically $J_H \gg t$.
Periodic boundary conditions (PBC) will be used. The model will be studied
with the Monte Carlo technique already 
described in Ref.~\cite{yuno}. This method basically amounts to
integrating exactly the $e_g$ electrons using library subroutines (since
it corresponds to a one-electron problem in the background of the
localized spins), supplemented by a Monte Carlo simulation
on the $t_{2g}$ classical spins. The technique
does not have ``sign'' problems  and it allows to reach low temperatures
at any density. The neglect of the direct Coulomb interactions among
the $e_g$ electrons can be justified recalling that a large Hund coupling
effectively reduces double occupancy, 
similarly as a large on-site Hubbard repulsion does.

\section{Phase Separation}

In Ref.\cite{yuno} the phase diagram of model Eq.(1) has been studied for the
special case 
$J'=0$. In that paper it was found that at low temperatures the
antiferromagnetic insulating phase that appears at $x=0$ is phase
separated from a metallic ferromagnetic phase that becomes stable for 
$x>0.25$. Here
the effect that a finite $J'$ has on this important result will be 
investigated.
 
To study phase separation the density $\langle n \rangle$ as a 
function of the chemical
potential $\mu$ has to be monitored. Working with a chain of $L$ sites
and at zero temperature ($T$),
the density $\langle n \rangle$ can take only the values ${m\over{L}}$ with $m$
running from 0 to $2L$. However, at finite $T$ the effects of thermal
fluctuations allow for intermediate values of $\langle n \rangle$ to exist, 
and this
complicates the study of phase separation since sharp discontinuities
are usually
transformed into rapid crossovers. This behavior is clearly observed in
long MC runs when tunneling between several of the allowed states
occurs. In particular during MC runs of 40,000 iterations at low temperature
the system was observed to tunnel several times between states with a
sharply defined value of
$\langle n \rangle$ for particular values of the chemical potential. 
In Fig.1 results for a Monte Carlo run for
$\mu=-6.97$ on an $L=20$ chain and at $T=1/75$ are presented as an
illustration. The density clearly
fluctuates between states with 15, 16, 17, 18 and 19 electrons.
This particular behavior allowed us to measure 
ground state properties for most of the possible filling levels
 by simply carrying out
measurements of, e.g., the density and energy in the plateaus observed in
Fig.1. With this
information (i.e. energy vs density without temperature contaminations)
 it was possible to
perform an accurate analytical calculation of $\langle n \rangle$ vs $\mu$ at
$T=0$ using the Maxwell construction. This
method allows us to obtain ground state information
from a finite temperature simulation.
\begin{figure}[htbp]
\centerline{\psfig{figure=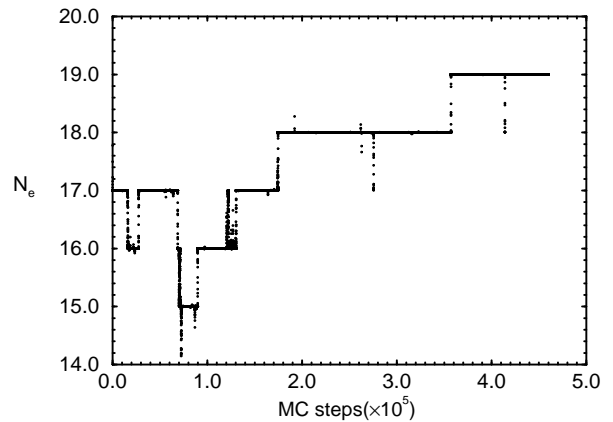,width=8.5cm,angle=-90}}
\vspace{0.5cm}
\caption{
The total number of itinerant electrons $N_e$ as a function of the
Monte Carlo steps for an $L=20$ chain with $J_H=8$, $J'=0.05$, $T=1/75$
and $\mu=-6.97$.
}
\vspace{0.5cm}
\end{figure}

In Fig.2a the relation $\langle n \rangle$ vs $\mu$ at $T=0$ obtained
analytically from the MC results at $J'=0.05$ on an $L=20$
lattice (solid line) is presented, and it is compared with the results
obtained directly
from Monte Carlo simulations at 
$T=0.01$ (open circles)~\cite{foot2}. The $T=0$ curve shows that one regime of
phase separation (denoted by
PS1) occurs between regions with $\langle n \rangle=1$ (i.e.,
$x=0$) and $\langle n \rangle=0.65$. 
This behavior is similar to the results recently reported in
Ref.\cite{yuno} working at $J'=0$. However, the same figure shows the novel
result that
there is a second region of phase
separation (denoted by PS2) in the limit of low $e_g$-electronic density 
between the empty system $\langle n \rangle=0.0$ 
and an electron-rich phase with $\langle n \rangle=0.25$.
Through a careful analysis it has been
found that phase separation at low density
also occurs for smaller values of $J'$ such as 0.02. 
\begin{figure}[htbp]
\centerline{\psfig{figure=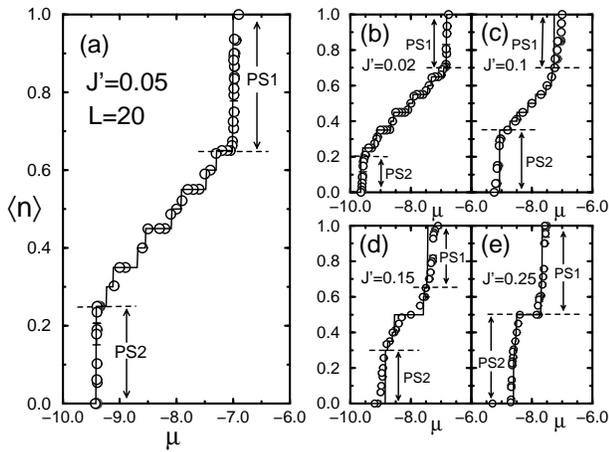,width=9.5cm,angle=-90}}
\vspace{0.5cm}
\caption{
The density of itinerant electrons $\langle n \rangle$ as a
function of the chemical potential $\mu$ on $L=20$ chains with $J_H=8$
and (a) $J'=0.05$, (b) $J'=0.02$, (c) $J'=0.1$, (d) $J'=0.15$ and (e)
$J'=0.25$. The open circles indicate Monte Carlo results at $T=0.01$.
The lines are obtained using the Maxwell construction and they indicate
$T=0$ behavior.
}
\vspace{0.5cm}
\end{figure}

It is important to remark that at $J'=0$ 
indications of a rapid crossover at low densities exist 
but this effect does not reach
the full character of 
phase separation since at $x=1$ the $t_{2g}$ spins are
non-interacting in this special case.
In Fig.2b-e, $\langle n \rangle$ vs $\mu$ curves at $T=0$ 
for different values of $J'$ using chains with
$L=20$ sites are presented (obtained using the methods discussed above
that allowed us to remove the temperature contaminations from the results). 
Additional MC runs using
open boundary conditions and different lattice sizes 
close to the phase
separation critical chemical potentials have shown that the results are not
affected by strong finite size effects, and thus we are confident that they
already represent the bulk limit. In general, it was observed (Fig.2)
that as $J'$ increases the range of stable densities decreases, and
for $J'\geq 0.25$ the only stable density besides $\langle n 
\rangle=1$ and 0 is $\langle n \rangle=0.5$.
Evenmore, for values of $J'$ larger than 0.6 only the empty and the
half-filled phases are stable. 
This remarkably strong tendency to phase separation can be qualitatively
understood in the limit where $J'$ is larger than $t$, both still smaller than
the Hund coupling. In this regime the antiferromagnetic order caused
by $J'$ among the localized spins induces a similar antiferromagnetic
tendency among the itinerant electrons (again, note that the Hund
coupling is assumed
to be the largest scale). Then, in the conduction band an effective 
one band $t-J'$ model with $J'/t \gg 1$ dominates the physics. It is
known that in this regime the $t-J'$ model phase separates
between an empty and half-filled phases in all dimensions~\cite{elbio}. 
This can also be understood easily by considering the atomic limit of
Eq.(1), i.e., when $t \ll J_H,J'$.
In this case the lowest energy of a state with $N_e$ $e_g$ electrons,
including the chemical potential, is
given by $-J_H N_e+J' L-\mu N_e$ if $N_e\leq L$ and 
$-J_H (2L-N_e)+J' L-\mu N_e$ if $N_e\geq L$. 
The system is half-filled for $\mu=0$. When $\mu=-J_H$ $(J_H)$ the energies
of all the
levels with $N_e\leq L$ ($N_e\geq L$) become degenerate with $E=J'L$
($(J'-2J_H)L)$ . 
For $\mu < -J_H$ ($\mu > J_H$) the state with
the lowest energy $E=J'L$ ($(J'-2J_H)L)$ is the one with $N_e=0$ ($2L$).
For $-J_H<\mu<J_H$ the ground state is the state with $N_e=L$. 
Thus, at $\mu=-J_H$ $(J_H)$ phase
separation between the half-filled and the empty (fully doubly occupied)
state occurs.
Similar
tendencies have been recently observed in one band models for 
manganites~\cite{riera}.

In Fig.3 the stable densities as a function of $J'$ are presented.
A very similar result was obtained using open boundary conditions.
In the following sections the properties of the stable
regions in the phase diagram will be studied. 
\begin{figure}[htbp]
\centerline{\psfig{figure=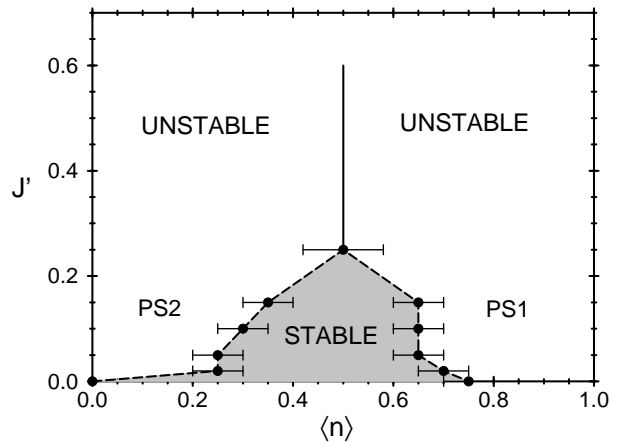,width=9cm,angle=-90}}
\vspace{0.5cm}
\caption{
The stable and unstable regions in the $J'$-$\langle n \rangle$
plane are indicated for $J_H=8$.
}
\vspace{0.5cm}
\end{figure}

\section{Magnetic and Charge Ordering}

\subsection{Magnetic}

Some of the most important properties of the manganite materials are
related to the magnetic order of the electrons since a remarkable
interplay between ferromagnetic and
antiferromagnetic phases has been observed in the
experiments~\cite{schi}. 
Actually recent neutron-scattering measurements performed 
on $\rm{La_{2-2x}Sr_{1+2x}Mn_2O_7}$ with
$x=0.4$ have reported the coexistence of short-range antiferromagnetic
correlations with strong ferromagnetic fluctuations~\cite{aep}. In
particular, antiferromagnetic fluctuations one order of magnitude smaller
in intensity
than the ferromagnetic ones were observed between 100K and 200K, i.e.,
above the Curie temperature. The antiferromagnetic fluctuations
disappear when ferromagnetic long-range order is established.

Motivated by these experiments, in the present section we will study the 
behavior of the spin-spin
correlation functions among the classical spins defined as
$$
\omega(r)={1\over{L}}\sum_i\langle{\bf{S}}_i\cdot{\bf{S}}_{i+r}\rangle,
\eqno(2)
$$
\noindent where the notation is standard.
The structure factor
$S(q)$ is given by the Fourier transform of the spin-spin correlations:
$$
S(q)=\sum_r e^{-iqr}\omega(r),
\eqno(3)
$$
\noindent where the momentum $q$ can take the values $2 \pi n/L$ with
$n$ running from 0 to $L-1$.
In Fig.4, $S(q)$ at $T=1/75$, $J'=0.05$ and for densities 
$\langle n \rangle=0$, 
0.35, 0.5, 0.65 and 1 is presented. It is
clearly observed in the figure that 
for $\langle n \rangle=0$ and 1 the structure factor peaks at momentum
$q=\pi$ denoting antiferromagnetism in the system in 
agreement with the experimental behavior~\cite{schi}. 
This is certainly the effect
that we wanted to induce by the inclusion of a $J'$ coupling, namely
the existence of AF correlations at $both$ ends of the $e_g$ density
range rather than only at $x=0$.
At the remaining stable
values of $\langle n \rangle$ the peak occurs at $q=0$ or
$q=2\pi/20$ signaling the presence of
ferromagnetism. We observed that when the number of itinerant electrons
is even and periodic boundary conditions are
used, a kink appears in the ground state separating two ferromagnetic
regions with opposite spin.\cite{kubo2} 
This kink causes the peak in $S(q)$ to move to $2
\pi/L$~\cite{foot3}.  As a result of this effect a kink appears at quarter
filling when the length of the chain $L$ is a multiple of 4. When this happens
we have verified that the peak is at $q=0$ for other values of $L$ 
and also using open and antiperiodic boundary conditions.
Then, there is no doubt that ferromagnetic correlations are dominant
when the peak in $S(q)$ occurs at or close to $q=0$.
\begin{figure}[htbp]
\centerline{\psfig{figure=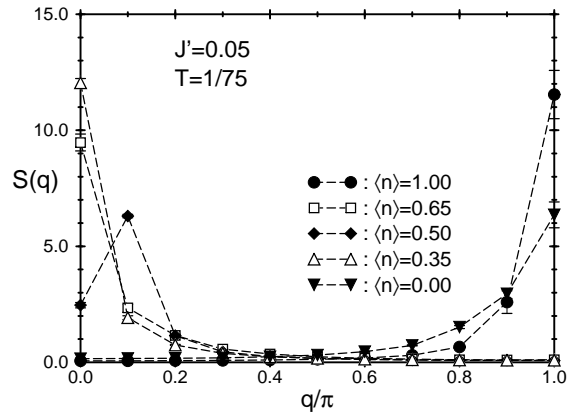,width=8.5cm,angle=-90}}
\vspace{0.5cm}
\caption{
Structure factor, $S(q)$, as a function of $q/\pi$ for an $L=20$
chain at $J'=0.05$ and $T=1/75$. The densities are indicated in the
figure.
}
\vspace{0.5cm}
\end{figure}
\begin{figure}[htbp]
\centerline{\psfig{figure=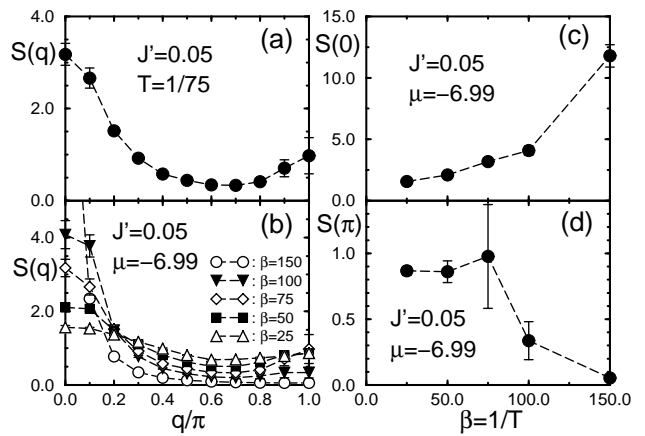,width=9cm,angle=-90}}
\vspace{0.5cm}
\caption{
(a) Structure factor as a function of $q/\pi$ on an $L=20$ chain at
$T=1/75$, $\mu=-6.99$ and $J'=0.05$. Maxima at $q=0$ and $\pi$ are
clearly observed; (b) same as (a) at different temperatures indicated in
the figure; (c) the structure factor at $q=0$ as a function of the
inverse temperature $\beta$ for $\mu=-6.99$; (d) same as (c) for $q=\pi$.
}
\vspace{0.5cm}
\end{figure}
 
Note that for all the densities studied in Fig.4 the structure factor
has a maximum
at a single value of the momentum indicating that only one kind of
magnetic fluctuations prevail at a given density. However, finite temperature 
measurements at other densities such
as $\langle n \rangle=0.75$ which corresponds to an {\it unstable} density
in the limit of zero temperature indicate otherwise.
As it can be observed in Fig.5a, strong 
ferromagnetic correlations coexist with weak antiferromagnetic ones at 
this density and $T=1/75$. 
This result has similarities with the experimental
data~\cite{aep} discussed before,
 and we believe it is due to the coexistence of hole-undoped
antiferromagnetic and
hole-rich ferromagnetic domains
rather than to a ferromagnetic spin polaron
in an antiferromagnetic background, as suggested in Ref.\cite{horsch}. To
explore this scenario the chemical potential was fixed at $\mu=-6.99$
and $S(q)$ was measured at $T=1/25$, 1/50, 1/75, 1/100 and 1/150. The results
are presented in Fig.5b. As the temperature is lowered the intensity of
the peak at $q=0$ increases (Fig.5c) while the peak at $q=\pi$
decreases (Fig.5d). The
density reaches the value 0.65 at the lowest temperature indicating that
the system at 0.75 is unstable. The qualitative similarities
of our results with the experimental data of Ref.~\cite{aep} is clear.
The relative intensity of the ferromagnetic and antiferromagnetic peaks
can be tuned by properly selecting the density at finite temperature
in the region that will become unstable as $T\rightarrow 0$.
\begin{figure}[htbp]
\centerline{\psfig{figure=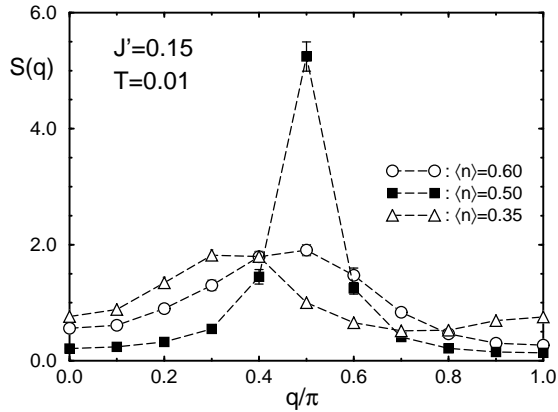,width=8.5cm,angle=-90}}
\vspace{0.5cm}
\caption{
Structure factor as a function of the momentum for different
densities at $J'=0.15$ and $T=0.01$.
}
\vspace{0.5cm}
\end{figure}

Ferromagnetism in the low-temperature stable intermediate densities between
the two regimes of phase separation PS1 and PS2
is observed up
to $J'\approx 0.1$. However, for larger values of $J'$ weak incommensurate
correlations were
observed. In Fig.6, $S(q)$ at $J'=0.15$ and for densities
 $\langle n \rangle= 0.35$, 0.5
and 0.6 is presented. 
A low-intensity peak at $\langle n \rangle=0.35$ and 0.6 
suggests a tendency to 
short-range incommensurate order with
momentum $q=0.3 \pi$ and $0.5 \pi$, respectively. 
Note, however, that at
 $\langle n \rangle=0.5$ a strong peak is observed at $q=\pi/2$. 
This indicates the
existence of a spin twisted ground state (a spiral state with pitch
$q=\pi/2$). The structure factor at
$\langle n \rangle=0.5$ in chains of sizes ranging 
from $L=10$ to 30 was studied 
and it is presented  in Fig.7a. 
A large increase of the peak intensity 
with the lattice size was not observed. However, in all the
chains studied we found that the
correlation length for the twisted order although finite seems fairly
large as it can be observed
in Fig.7b.
The intensity of the incommensurate correlations at
quarter-filling reaches a maximum at $J'\approx 0.15$. As $J'$ increases
further,
incommensurability at $q=\pi/2$ weakens and at $J' \approx 0.35$ the
ground state becomes antiferromagnetic, which is  the expected behavior at
the stable densities in the large $J'$ limit.
\begin{figure}[htbp]
\centerline{\psfig{figure=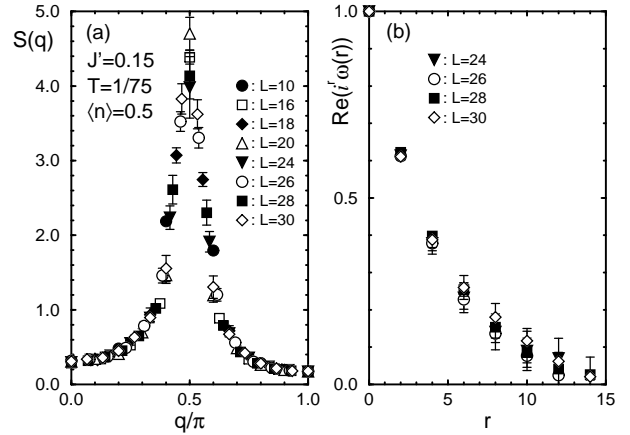,width=8.5cm,angle=-90}}
\vspace{0.5cm}
\caption{
(a) Structure factor as a function of $q/\pi$ at $T=1/75$,
$J'=0.15$ and $\langle n \rangle=0.5$ for L=10, 16, 18, 20, 24, 26, 28
and 30. A peak at $q=\pi/2$ appears; (b) real part of the spin
correlation function $\omega(r)$ multiplied by $i^r$ as a function of
the distance $r$ for $L=24$ (triangles), 26 (circles)
$L=28$ (squares) and 30 (diamonds).
}
\vspace{0.5cm}
\end{figure}

\subsection{ Charge}

The existence of commensurate and incommensurate charge ordered phases has been
reported in the experimental 
literature for $\rm{La_{1-x}Ca_xMnO_3}$~\cite{chen}. It was
observed that for values of $0.5\leq x \leq 0.67$ antiferromagnetism coexists
with commensurate charge order, and charge incommensurability was detected
in a ferromagnetic phase at higher temperature. This behavior is very
puzzling since ferromagnetism and charge ordering were not 
expected to appear together, particularly based on analysis using 
the double exchange model.
In order to analyze the effect of the interplay of double
exchange and antiferromagnetic interactions in charge ordering
for the itinerant fermions, the
Fourier transform $N(q)$ of the charge correlations was studied. 
The charge correlation functions are given by
$$
n(r)={1\over{L}}\sum_i \langle (n_i-\langle n \rangle)( n_{i+r}-\langle
n \rangle)\rangle ,
\eqno(4)
$$
\noindent where $n_i$ is the number operator at site $i$
 for the itinerant fermions.
Computationally we
 found that in the antiferromagnetic phase at half-filling $N(q)$ peaks
at $q=\pi$ rather than $q=0$. This behavior is due to  charge 
correlations that are negative (charge repulsion) 
at very short distances rather than to long range ordering, and it is
in agreement with previous calculations~\cite{chino}. 
In the ferromagnetic
phase we observed that $N(q)$ behaves as for non-interacting
spinless fermions, i.e. it reaches a maximum 
at $q=2k^{sf}_F=2 \pi \langle n \rangle$ ($k^{sf}_F=\pi \langle n
\rangle$ is the Fermi momentum for free spinless fermions) 
as $q$ grows from 0, and
then it becomes flat (Fig.8-a). This indicates that the value of the Hund
coupling, $J_H=8$, is sufficiently 
strong for the itinerant spins to be parallel to
the localized ones, behaving effectively as 
non-interacting spinless fermions in the charge channel.
\begin{figure}[htbp]
\centerline{\psfig{figure=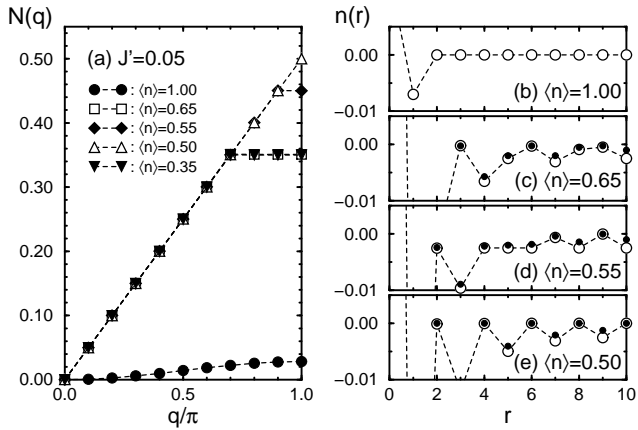,width=8.5cm,angle=-90}}
\vspace{0.5cm}
\caption{
(a) Fourier transform of the charge correlation functions as a
function of $q/\pi$ at the densities indicated in the figure for
$J'=0.05$ on a $L=20$ chain at $T=1/75$; (b) Charge correlation function
as a function of distance for $\langle n \rangle=1$; (c) same as (b) for
$\langle n \rangle=0.65$. The small full circles are results obtained
using Eq.(5); (d) same as (c) for $\langle n \rangle=0.55$; (e) same as (c)
for $\langle n \rangle=0.5$.
}
\vspace{0.5cm}
\end{figure}

In Fig.8b-e the charge correlations as a function of distance are
presented. At half-filling (Fig.8b) the only non-negligible
correlations are the nearest-neighbor ones which
are negative. This effect induces a maximum at $q=\pi$ in $N(q)$. In the
ferromagnetic region the charge correlations show oscillatory behavior.
In particular, at quarter filling (Fig.8e) the correlations are zero at
even distances and negative at odd ones. Although the system is
metallic, this produces a maximum of
$N(q)$ at $q=\pi$ similar to the one experimentally
observed along the $a$ axis of Ca-doped manganites with
$50$\% doping~\cite{chen}. The behavior of the charge correlation for
non-interacting spinless fermions is given by
$$
n(r)={-1\over{2 \pi^2 r^2}}+{\cos(2k^{sf}_Fr)\over{2 \pi^2 r^2}}.
\eqno(5)
$$ 
This expression is strictly valid in the limit $L\rightarrow\infty$ and
for $r \neq 0$ but
as it can be observed from 
Fig.8c, d and e, where the correlations obtained from Eq.(5) are
denoted with filled circles, they are in very good agreement with the numerical
results for $L=20$.
Thus, apparently
this is a system that behaves like a ferromagnet in the spin
channel, but is non-interacting in the charge channel.

While the qualitative behavior of $N(q)$ at half-filling is independent 
of $J'$ since the ground state is always antiferromagnetic, a
different behavior is observed at intermediate densities when the ground
state is no longer ferromagnetic. 
In Fig.9a, $N(q)$ is presented for $J'=0.15$. $N(q)$ now reaches a maximum at
$q=2k^{sf}_F$ and afterwards the intensity forms a plateau at a lower value. In
Fig.9b-e the charge correlations as a function of the distance are
shown. It is clear that the peaks in $N(q)$ are caused by oscillatory
behavior of the charge correlations. It appears that a tendency to 
short-range charge
ordering exists in the model for densities close to quarter-filling. 
\begin{figure}[htbp]
\centerline{\psfig{figure=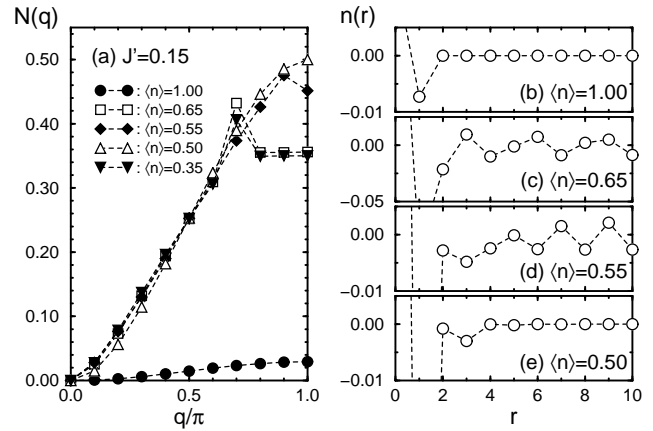,width=8.5cm,angle=-90}}
\vspace{0.5cm}
\caption{
(a) Fourier transform of the charge correlation functions as a
function of $q/\pi$ at the densities indicated in the figure for
$J'=0.15$ on a $L=20$ chain at $T=0.01$; (b) Charge correlation function
as a function of distance for $\langle n \rangle=1$; (c) same as (b) for
$\langle n \rangle=0.65$; (d) same as (c) for $\langle n \rangle=0.55$;
(e) same as (c) for $\langle n \rangle=0.5$.
}
\vspace{0.5cm}
\end{figure}

\section{Fermi Surface and Spectral Functions}

In previous sections, regions in parameter space with a variety of
magnetic and charge structures were found. Here the
dynamical properties of these phases will be studied. A calculation of the
spectral function $A(k,\omega)$ and the study of the momentum
distribution function 
$n(k)$ of the $e_g$ electrons 
will allow us to determine, at least qualitatively,
 whether a Fermi surface exists and,
thus, whether the phase is insulating or metallic (this analysis will 
continue in the following section where the Drude weight of the optical
conductivity
will be discussed). In principle, 
the spectral function can be compared with angle-resolved
photoemission (ARPES) measurements performed in
$\rm{La_{1.2}Sr_{1.8}Mn_2O_7}$\cite{dessau} and in 
$\rm{La_{1-x}Sr_xMnO_3}$~\cite{saitoh}.

The spectral function $A(k,\omega)$ is defined as
$$
A(k,\omega)=-{1\over{\pi}}Im G(k,\omega),
\eqno(6)
$$
\noindent where $G(k,\omega)$ is the one-particle
Green's function for the $e_g$
electrons. Since we are performing a Monte Carlo calculation
on the classical spins only, the time-dependent Green's function can be
straightforwardly calculated in real-time since the fermions do not 
interact among themselves directly, but only with the classical spins.
The spectral functions 
obtained with this procedure
only contain small controlled statistical errors. This should be
contrasted against other calculations that use the usually uncontrolled
Maximum Entropy technique.

Studying the density of states for the Hamiltonian defined in Eq.(1) we
observed low-energy and high-energy bands. The separation between
these two bands is regulated by the Hund coupling $J_H$, i.e., the
lower (upper)-band includes states where the spin of one of the
itinerant electrons
at an arbitrary site $i$ is parallel
(antiparallel) to the classical spin at the same site of the lattice. The
gap is about $2J_H$. 
Each of the two bands has an internal structure and a
bandwidth which are regulated in part by the antiferromagnetic coupling $J'$.
\vspace{0.7cm}

\subsection{Half-filling}
\vspace{0.5cm}

In Fig.10a the density of states $N(\omega)$ corresponding to
 $J'=0$, half-filling,
and with $T=1/100$ on a L=20 chain 
is presented.\cite{foot4} The bandwidth of the upper and lower
bands is $\sim 1.9t$ and the gap separating them is $\sim 14t$. 
The chemical potential
lies in the gap indicating that the system is an insulator.
In Fig.10b, $A(k,\omega)$ is shown. 
Both  below and
above the chemical potential,
spectral weight is observed
for {\it all} of the momenta $k$,
rather than only for $k$ below or above the Fermi momentum as in
non-interacting electronic systems.
This is reminiscent of the antiferromagnetically induced features
observed at half-filling in studies of the one band Hubbard model
for cuprates~\cite{haas}.
Below the
chemical potential there is a sharp  peak for $k$
smaller than $\pi/2$ at the bottom of the lower band.
The dispersion of this feature
is about $t$. The corresponding opposite behavior is observed in
the upper band.
If $J'$ is increased, the bandwidth of the upper and lower bands
is slightly reduced due to the increase in the AF correlations, 
as it can be observed in Fig.11 where the half-filled density of
states is presented for $J'=0$, 0.05, 0.1, 0.15 and 0.25. 
\begin{figure}[htbp]
\centerline{\psfig{figure=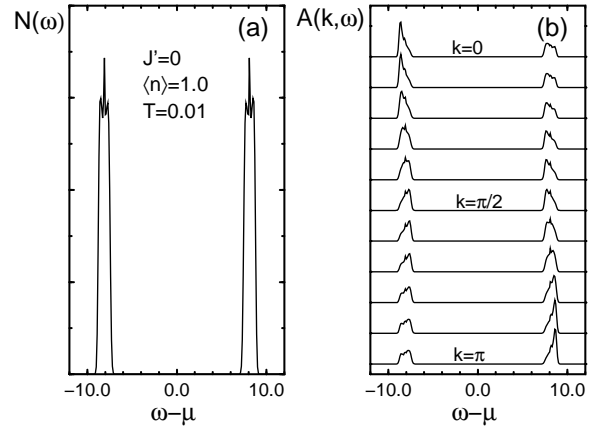,width=9cm,angle=-90}}
\vspace{0.5cm}
\caption{
(a) Density of sates $N(\omega)$ as a function of $\omega-\mu$ for
$J_H=8$, $J'=0$ at $\langle n \rangle=1$ for $L=20$ at $T=0.01$; (b) the
spectral function $A(k,\omega)$ as a function of $\omega-\mu$ for the
values of $k$ available on an $L=20$ lattice. The parameters are as in
(a).
}
\end{figure}
\vspace{-0.9cm}
\begin{figure}[htbp]
\centerline{\psfig{figure=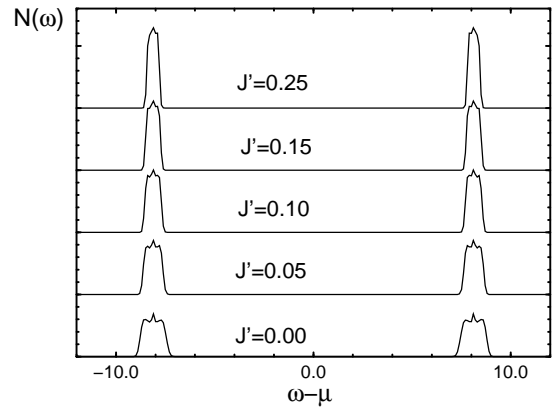,width=8.5cm,angle=-90}}
\vspace{0.5cm}
\caption{
Density of states as a function of $\omega-\mu$ at half-filling on
$L=20$ chains at $T=0.01$ for the values of $J'$ indicated in the
figure.
}
\vspace{0.5cm}
\end{figure}

\subsection{Quarter-filling}

The spectral functions change drastically in the
ferromagnetic phase. First let us discuss results
for density $\langle n \rangle = 0.5$ (quarter-filled).
$J_H$ still regulates the gap between the upper and lower bands. 
We will concentrate our attention on the lower band, since 
symmetrical features are observed in the upper band.
In Fig.12a, $A(k,\omega)$
at $J'=0$ is shown at quarter-filling and $T=1/75$ on a $L=40$ chain
(similar results were obtained for shorter chains). 
The chemical potential 
lies now in the middle of the lower band 
indicating that the system is a metal with a Fermi surface at
$k_F=\pi/2$. Notice that $k_F$ results to be equal to $k^{sf}_F$. 
For each
momentum $k$ there is a sharp peak which is naturally associated
with a quasiparticle peak. The dispersion resembles results
for a system of free spinless fermions (indicated by a dashed line in
the figure), in agreement with
the behavior of the charge correlations described in the previous
section. The bandwidth
is $4t$. The density of states can be seen in Fig.12b.
It resembles $N(\omega)$ for spinless fermions at half-filling which
for a finite length chain
typically presents several spikes that become a smooth function only
for large enough chains. Thus, the spikes observed in Fig.12b near
the chemical potential are likely finite size effects, and the
quasiparticle peak crosses the chemical potential as in a standard metal.
Similar behavior is observed for $J'$ up to approximately 0.11 in the
ferromagnetic phase. 
\begin{figure}[htbp]
\centerline{\psfig{figure=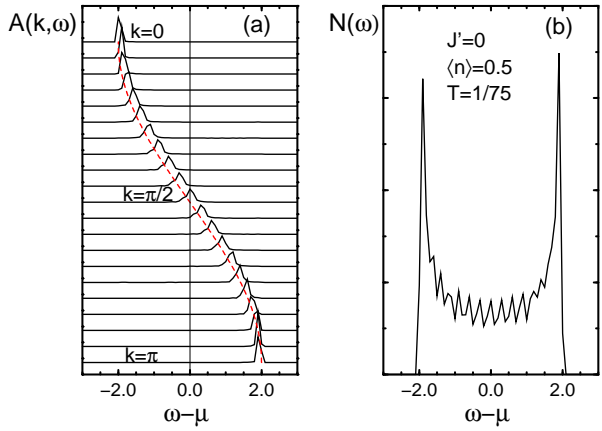,width=9cm,angle=-90}}
\vspace{0.5cm}
\caption{
(a) The spectral function $A(k,\omega)$ as a function of
$\omega-\mu$ for the values of $k$ available on an $L=40$ lattice for
$J_H=8$, $J'=0$, $T=1/75$ at $\langle n \rangle=0.5$. The dashed line is
the dispersion for spinless fermions; (b) Density of
states as a function of $\omega-\mu$. The parameters are as in (a).
}
\vspace{0.5cm}
\end{figure}

However, as $J'$ increases and the incommensurate phase is
reached the behavior of the spectral function at quarter-filling changes
dramatically. Results for $J'=0.15$ are presented in Fig.13 (remember
that for this value of $J'$ strong
incommensurate spin correlations were detected in the study of $S(q)$).
A gap has now
opened and the chemical potential lies in it indicating that the system
became an insulator. As shown in Fig.13-b the size of the gap appears to
be independent of the size of the chains. 
Fig.13-a shows that considerable spectral weight appears above and below the 
chemical potential for momenta close to the Fermi
momentum $k_F^{sf}=\pi/2$. The dispersion for the features above and below
the chemical potential in the lower band is of the order of 2.5t.
With increasing $J'$ the system remains insulating, 
but it was observed that the
size of the gap decreases while antiferromagnetic correlations increase.
The gap closes up for $J' \approx 0.6$ when the quarter-filled state is
no longer stable. In Fig.14 the density of states
is presented for
different values of $J'$ at $\langle n \rangle=0.5$ using $L=20$ chains.
Once again, the spiky structure for $J' = 0.1$ or smaller is a finite
size effect that disappears as the size of the lattice grows. However,
the gap observed for $J' \ge 0.11$ is not spurious and signals an
insulating regime. The opening of this gap is not just a one dimensional
effect. In Fig.15 we present the density of states at quarter-filling
for $J_H=8$ and $J'=0.25$ at $T=1/50$ on a $4\times 4$ and a $6 \times
6$ lattice with PBC. It can be seen that a gap has opened in the lower band and
the chemical potential lies in it, denoting insulating behavior. The
finite size effects do not seem strong. 
\begin{figure}[htbp]
\centerline{\psfig{figure=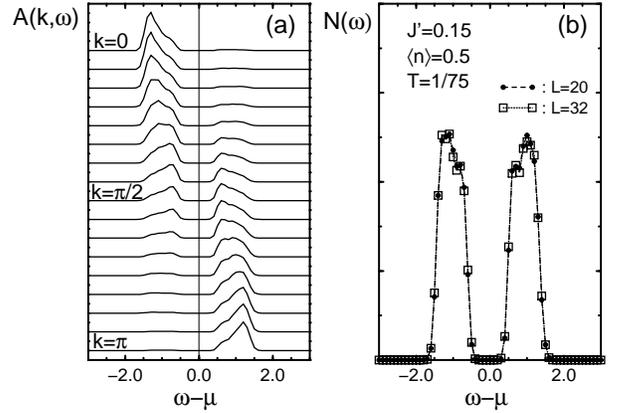,width=8.7cm,angle=-90}}
\vspace{0.5cm}
\caption{
(a) The spectral function $A(k,\omega)$ as a function of
$\omega-\mu$ for the values of $k$ available on an $L=20$ lattice for
$J_H=8$, $J'=0.15$, $T=1/75$ at $\langle n \rangle=0.5$; (b) Density of
states as a function of $\omega-\mu$. The parameters are as in (a) but
$L=20$ and 32.
}
\end{figure}
\vspace{-0.9cm}
\begin{figure}[htbp]
\centerline{\psfig{figure=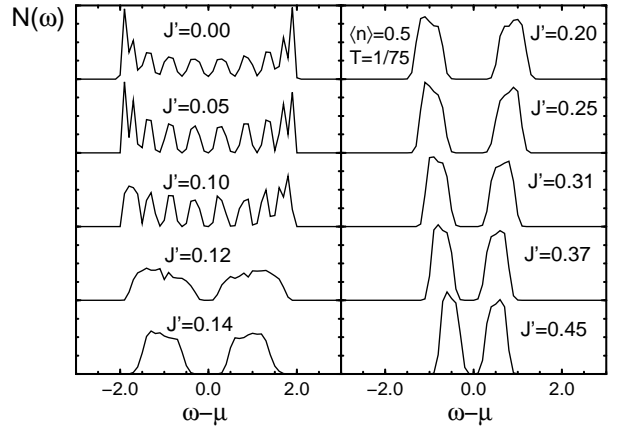,width=9cm,angle=-90}}
\vspace{0.5cm}
\caption{
Density of states as a function of $\omega-\mu$ for different
values of $J'$ at quarter filling for $L=20$ and $T=1/75$.
}
\end{figure}
\vspace{0.5cm}
\begin{figure}[htbp]
\centerline{\psfig{figure=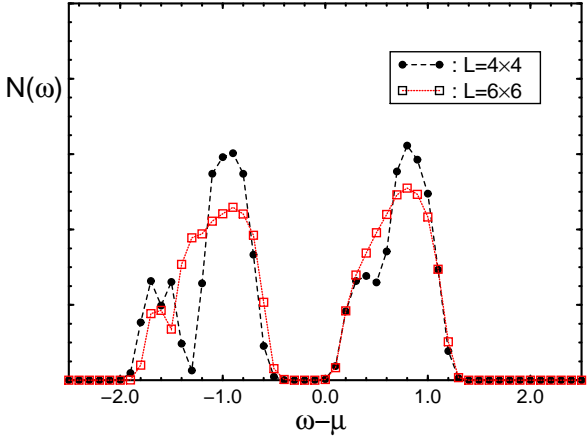,width=9.2cm,angle=-90}}
\vspace{0.5cm}
\caption{
Density of states as function of $\omega-\mu$ for $J'=0.25$ at
quarter filling for $4 \times 4$ and $6\times 6$ lattices at $T=1/50$
with PBC.
}
\vspace{0.5cm}
\end{figure}

\subsection{Other Densities}
\vspace{0.5cm}

The spectral functions at densities below and above 0.5 were also studied.
In the ferromagnetic region a behavior resembling
free spinless fermions was
observed, as it is shown in
Fig.16 where $A(k,\omega)$ for $J'=0.1$ is presented at (a)
$\langle n \rangle=0.55$
and (b) $\langle n \rangle=0.45$. As pointed out in Sec.IV.B, in the
charge channel the itinerant electrons behave as free spinless fermions
with $k^{sf}_F=\pi \langle n \rangle$. 
In the insulating phase, on the other
hand, the density of states changes qualitatively away from quarter
filling. When $N_e$ electrons ($N_h$ holes) are introduced in the
quarter filled state, spectral weight is redistributed so that a new
peak develops inside the gap and, thus, a three peak
structure is observed. The spectral weight in the original two peaks
changes from $L/2$ to $L/2-N_e$ ($L/2-N_h$) while the spectral weight of
the new structure is given by $2N_e$ ($2N_h$).\cite{spectral}  
However, the chemical potential still lies in the
gap indicating that the system continues to be an insulator as it can be
seen in Fig.17 where the density of states for $J'=0.15$ at different
values of $\langle n \rangle$ is presented. For the cases of $\langle n
\rangle=0.55$ and 0.45 the spectral function is shown in Fig.18. It can
clearly be seen that, for all values of the momentum $k$, spectral weight
is being transferred to the middle of the gap. Notice that while the two
external features disperse as a function of the momentum $k$, the peak
at the center does not. This third peak is located exactly in the middle
of the gap and it indicates that as electrons or holes are added to the
quarter filled insulating system, localized states form in the middle of
the gap.
\begin{figure}[htbp]
\centerline{\psfig{figure=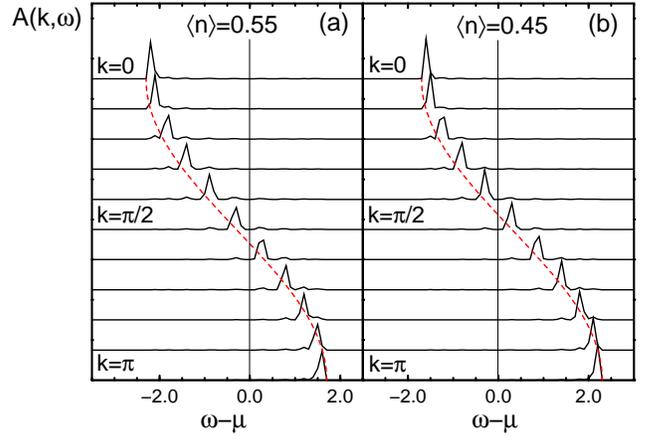,width=9.2cm,angle=-90}}
\vspace{0.5cm}
\caption{
(a) The spectral function $A(k,\omega)$ as a function of
$\omega-\mu$ for the values of $k$ available on an $L=20$ lattice for
$J_H=8$, $J'=0.1$, $T=1/100$ at $\langle n \rangle=0.55$; (b) same as (a)
for $\langle n \rangle=0.45$. The dashed line is the dispersion for
spinless fermions.
}
\vspace{0.5cm}
\end{figure}
\begin{figure}[htbp]
\centerline{\psfig{figure=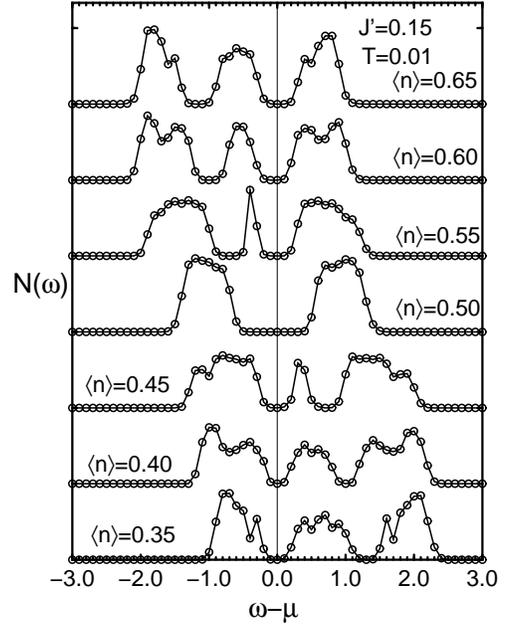,height=10cm,angle=0}}
\vspace{0.5cm}
\caption{
Density of states $N(\omega)$ as a function of $\omega-\mu$ for
$J_H=8$, $J'=0.15$ at different densities for $L=20$ at $T=0.01$.
}
\vspace{0.5cm}
\end{figure}
\begin{figure}[htbp]
\centerline{\psfig{figure=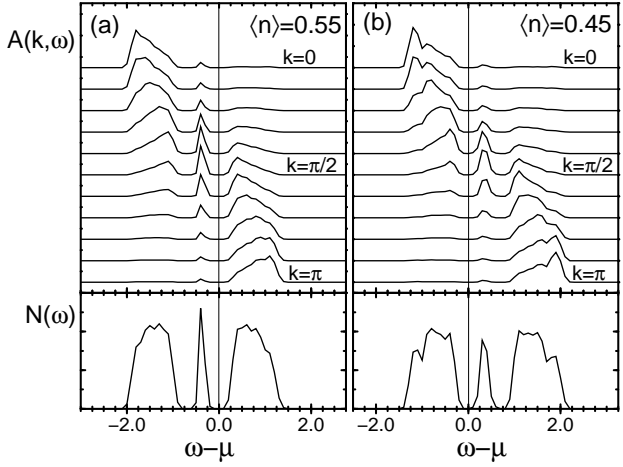,width=9cm,angle=-90}}
\vspace{0.5cm}
\caption{
(a) The spectral function $A(k,\omega)$ and the density of
states $N(\omega)$ as a function of $\omega-\mu$ on an $L=20$ lattice for
$J_H=8$, $J'=0.15$ at $\langle n \rangle=0.55$ at $T=0.01$; (b)
same as (a) for $\langle n \rangle=0.45$.
}
\vspace{0.5cm}
\end{figure}

The occupation number $n(k)$ can be calculated directly from the Green's
functions or just by integrating $A(k,\omega)$ on $\omega$ up to the
chemical potential. In Fig.19  $n(k)$ for quarter filling and
different values of $J'$ is presented. A sharp change in $n(k)$ suggests the
existence of a Fermi surface while a smooth behavior is more compatible
with the
opening of a gap. The change of behavior between $J'=0.10$ and 0.15 is
very clear. The analysis of this curves complements the previous
study of the spectral functions. Analogous results were
 obtained in both cases.
\begin{figure}[htbp]
\centerline{\psfig{figure=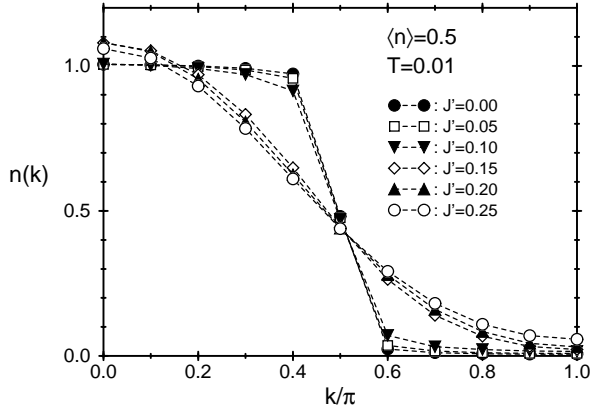,width=8.5cm,angle=-90}}
\vspace{0.5cm}
\caption{
$n(k)$ as a function of $k$ for different values of $J'$ for
$T=0.01$ at quarter filling on an $L=20$ chain.
}
\vspace{0.5cm}
\end{figure}

Some experimental
 ARPES studies for manganites
have found a strong dependence of the density of
states with the temperature~\cite{dessau}. In particular the appearance
of a pseudogap with increasing temperature was remarked. In Fig.20 the
density of states at quarter-filling for $J'=0$ is presented at
different temperatures. As the
temperature increases the peaks due to finite size effects are smeared
out but the density of states retains its qualitative characteristics.
There are no indications of a pseudogap developing as the temperature
increases. Thus, it appears that some
experimental observations may be due to effects that go
beyond the electronic interactions considered in this work. Probably
more than one orbital per site as well as phonons are needed to
account for this nontrivial effect.
\begin{figure}[htbp]
\centerline{\psfig{figure=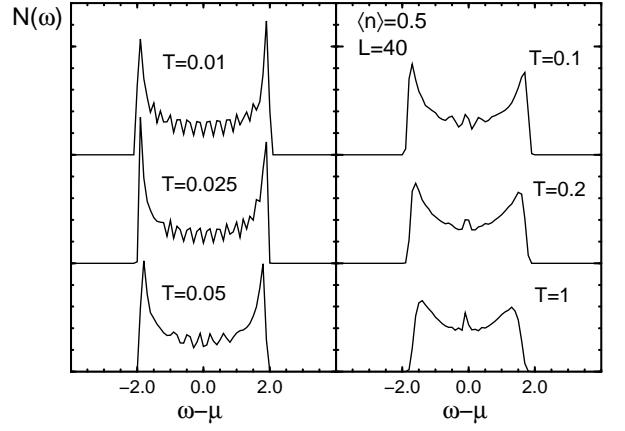,width=9cm,angle=-90}}
\vspace{0.5cm}
\caption{
The density of states as a function of $\omega-\mu$ at different
temperatures for $J_H=8$, $J'=0$ at quarter filling on a 40 site chain.
}
\end{figure}
\vspace{-0.5cm}
\begin{figure}[htbp]
\centerline{\psfig{figure=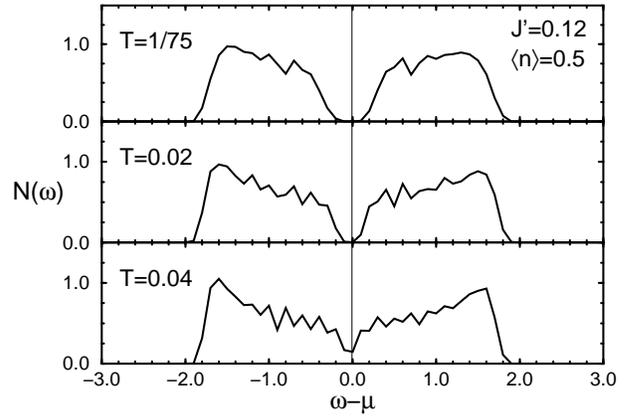,width=9cm,angle=-90}}
\vspace{0.5cm}
\caption{
The density of states as a function of $\omega-\mu$ at different
temperatures for $J_H=8$, $J'=0.12$ at quarter filling on a 22 site chain.
}
\vspace{0.5cm}
\end{figure}

For completeness, the temperature dependence of the density of states was
also studied at quarter-filling in the insulating phase. In Fig.21, where
results for $J'=0.12$ are presented, it can be seen that an increase in
temperature closes the gap in the density of states transforming it into
a pseudogap. These results suggest 
that the real manganite materials could be located
right at the border between the metallic and insulating regimes reported
here.


\section{Optical Conductivity and Drude Weight}

A very important characteristic of the manganite materials
 is the large change in
resistivity observed experimentally with small changes in
 doping, temperature, or magnetic fields. In addition, 
experimental measurements
of the optical conductivity in $\rm{La_{1-x}Sr_xMnO_3}$ with 
$0 \leq x \leq 0.3$~\cite{okimoto} indicate the existence of a gap in the
optical spectrum at half-filling which becomes a pseudogap in the
ferromagnetic region at high temperatures. 
This behavior has  similarities with those 
observed in ARPES data.
To study the extent to which electronic interactions can
account for the
optical measurements, here we have
calculated the optical conductivity defined as:
$$
\sigma(\omega)={\pi(1-e^{-\beta\omega})\over{\omega L}}\times
\int_{-\infty}^{\infty}{dt\over{2 \pi}}
e^{i \omega t} \langle j_x(t)j_x(0)\rangle,
\eqno(7)
$$
\noindent where $j_x$ is the current operator 
in the $x$-direction given by
$$
j_x=ite\sum_{j,s}(c^\dagger_{j+\hat x, s} c_{j, s} - h.c.),
\eqno(8)
$$
\noindent with the rest of the notation standard.
At $\omega=0$ there is a Drude weight with an intensity $D$ given by:
$$
{D\over{2 \pi e^2}}={<- \hat T>\over{4L}}-{1\over{2 \pi
e^2}}\int^{\infty}_{0^+} d \omega \sigma(\omega),
\eqno(9)
$$
\noindent where $\hat T$ is the kinetic energy operator and $e$ is the
electronic charge. Hereafter we set $e=1$.

In Fig.22a the Drude weight at density $\langle n \rangle=0.5$ 
is presented as a function of
$J'$ for chains of different lengths, and at low temperature. 
The results suggest that the Drude weight of the system decreases as $J'$
increases and it actually vanishes for $J' \approx 0.11$, i.e., where the
metal-insulator transition was observed in previous sections
by monitoring the spectral
function and $n(k)$. 
The effect of the temperature on the optical conductivity was also
studied. In Fig.22b the Drude weight is plotted as a function of      
temperature for several values of the antiferromagnetic coupling $J'$ 
working at quarter-filling.
For values of $J'$ that lead to a ferromagnetic 
ground state at low temperature,
the Drude weight increases as the temperature decreases. On the other hand,
a decrease in the Drude weight with decreasing temperatures
 is observed when the ground state at low
temperature is an insulator with strong spin-incommensurate correlations.
\begin{figure}[htbp]
\centerline{\psfig{figure=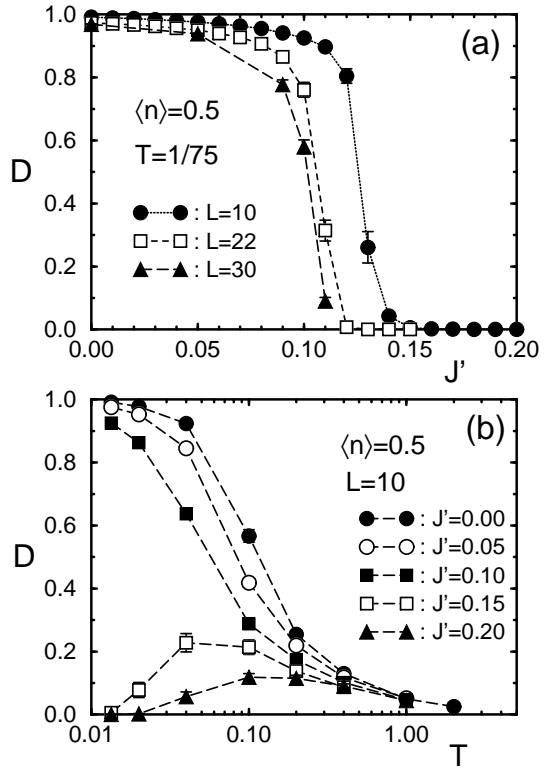,height=11cm,angle=0}}
\vspace{0.5cm}
\caption{
(a) The Drude weight for $\langle n\rangle=0.5$ as a function of
$J'$ for chains of different lengths; (b) Drude weight as a function of
temperature at quarter filling. The values of $J'$ are indicated in the
figure.
}
\vspace{0.5cm}
\end{figure}

In Fig.23 the optical conductivity for $J'=0$ at different temperatures
is presented. It is clear that the spectral
weight on the high-energy side, i.e. across the $2J_H$ gap, decreases as the
temperature decreases while the spectral weight on the low lying band
increases. This occurs because at high temperature both the spins up and
down have similar occupations of their lower bands and the optical gap
transitions have contributions from both kind of spins. As the
temperature is lowered and ferromagnetism develops, the density of
states for the spins antiparallel to the classical spins goes up in
energy while the density of states for the parallel spins at low energy
increases providing low energy states for transitions, and as a consequence
the transitions
across the $2J_H$ gap  diminish. Similar results have been obtained in
infinite dimension.\cite{furu2}
This behavior is in qualitative
agreement with experimental measurements of the optical conductivity in
doped $\rm{LaMnO_3}$\cite{okimoto}.
\begin{figure}[htbp]
\centerline{\psfig{figure=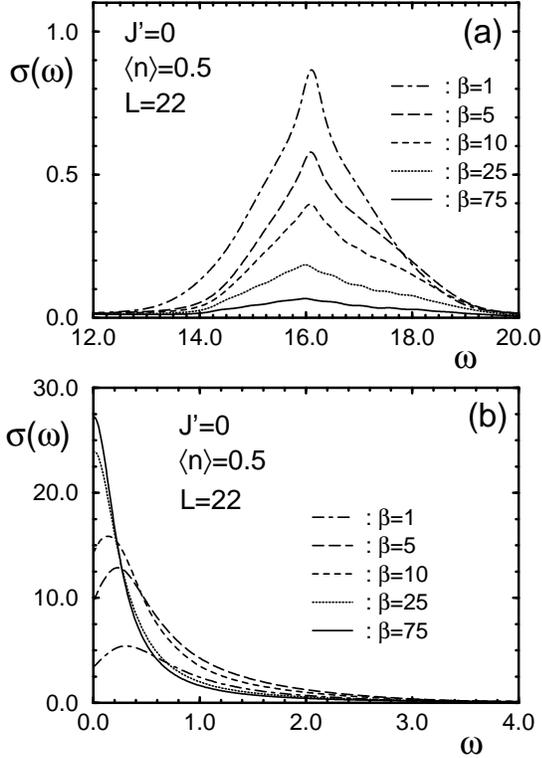,height=11cm,angle=0}}
\vspace{0.5cm}
\caption{
Optical conductivity as a function of $\omega$ for $J_H=8$,
$J'=0$ and quarter filling for different values of the temperature; (a)
high energy; (b) low energy. A lorentzian with width 0.25 was used.
}
\end{figure}
\begin{figure}[htbp]
\centerline{\psfig{figure=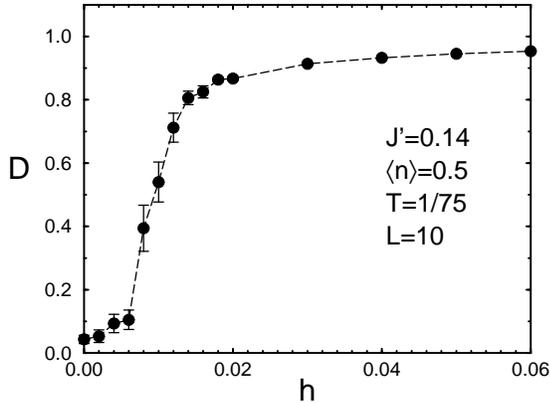,width=8.5cm,angle=-90}}
\vspace{0.3cm}
\caption{
Drude weight as a function of magnetic field for $J'=0.14$,
$T=1/75$ at quarter filling on an $L=10$ chain.
}
\vspace{0.5cm}
\end{figure}

Finally, for completeness the influence 
of a magnetic field on our results was also considered. The external
magnetic field was coupled to the itinerant and localized electrons via
a Zeeman term, $-h
\sum_i(\sum_{\alpha,\beta}c^{\dagger}_{i,\alpha}(\sigma_z)_{\alpha,\beta}
c_{i,\beta}+3S_i^z/2)$ .
A insulator-metal transition induced by the magnetic field $h$ was observed
for values of $J'$ where the ground state is an insulator, as it can be
seen in Fig.24. 

\section{Phase Diagram}

The information about the Kondo Hamiltonian presented in the previous
sections allows us to obtain the low temperature phase diagram as a
function of the antiferromagnetic coupling $J'$ and the density $\langle
n \rangle$ in the limit of large Hund coupling.
The full phase diagram is presented in Fig.25. For $J'$ smaller than
$\sim 0.11$ the
ground state is a ferromagnetic metal for $\langle n \rangle$ 
between $\sim 0.25$ and $\sim 0.65$ and
an antiferromagnetic insulator for the empty and half-filled systems
that are phase separated from the intermediate densities. This is the
region of the phase diagram which is relevant for comparisons with
experimental results for the manganites. When $J'$ becomes larger than
$\sim 0.11$ a metal-insulator transition occurs and the system becomes an
incommensurate insulator at intermediate fillings. In particular the
ground state at quarter-filling appears to be a spin  spiral state with a
maximum at $q=\pi/2$ in the magnetic structure factor, and finite but very long
correlation lengths. The allowed intermediate densities diminish as
$J'$ increases and for $J' \approx 0.25$ only quarter-filling is stable.
At $J' \approx 0.35$ the quarter filled ground state becomes
antiferromagnetic and
for $J'$ larger than $\sim 0.6$ only the half-filled and the empty phases
are stable. Both of them correspond to antiferromagnetic insulators.
\begin{figure}[htbp]
\centerline{\psfig{figure=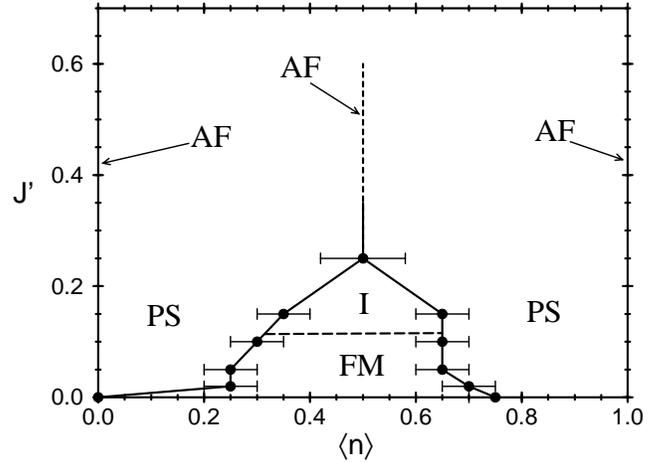,width=9.5cm,angle=-90}}
\vspace{0.5cm}
\caption{
The phase diagram for the Kondo lattice Hamiltonian with
ferromagnetic Hund's coupling $J_H=8$. AF indicates the
antiferromagnetic phases, I the insulating region, FM
the ferromagnetic phase and PS denotes phase separation.
}
\vspace{0.5cm}
\end{figure}

\section{Summary}

In this paper we have studied a ferromagnetic Kondo lattice 
model with classical
localized spins coupled antiferromagnetically, using Monte Carlo methods. 
Working at large Hund coupling we found a very rich phase diagram as the density of itinerant electrons
and the strength of the antiferromagnetic coupling $J'$ was changed. 
In particular, phase separation between hole-undoped 
antiferromagnetic and hole-rich ferromagnetic states was observed at
high electronic densities. Reciprocally, phase separation between
electron-undoped antiferromagnetic and electron-rich ferromagnetic
regions was observed in the limit of low $e_g$ electronic density.
At intermediate densities, we found a
metallic ferromagnetic phase for small $J'$ separated from an insulating
regime with a 
transition located at  $J'\approx 0.11$. The insulating phase has very strong
incommensurate magnetic correlations with a structure factor that peaks
at $q=\pi/2$ for quarter-filling. 
At $J'\approx 0.35$ the quarter-filled ground state becomes
antiferromagnetic and beyond $J'\approx 0.6$ only the half-filled and
empty antiferromagnetic phases are stable.

$J'=0.05$ is possibly a good value to compare with experimental results.
We have observed ferromagnetic and antiferromagnetic spin correlations
coexisting for $\langle n \rangle=0.75$ at $T=1/75$. As in neutron
scattering experiments the ferromagnetic peak increases as the
temperature decreases, while the antiferromagnetic excitations vanish
in the ferromagnetic phase.
In the present model this behavior is due to the coexistence of
ferromagnetic and antiferromagnetic domains at high temperature,
which are a consequence of the tendency of the system to phase
separate~\cite{yuno}. 

The study of dynamical properties shows the existence of interesting
metal-insulator
transitions as a function of temperature and $J'$. A transference of
spectral weight from higher to lower energies with decreasing
temperature, in agreement with experimental results, is observed in the 
optical conductivity. The metal-insulator transition with increasing
$J'$ is a feature of the model also present in two dimensions and it is
very likely that the transition will also occur in three dimensions.

Summarizing, the present results show that several properties of the
manganites, such as the change in magnetic ordering as a function of
concentration and transference of spectral weight from high to 
low energies in the optical conductivity with temperature, may
be attributed principally to simple electronic interactions. However, additional
factors will have to be considered to explain, for example, the
opening of gaps in the density of states with increasing temperature.
The addition of phononic degrees of freedom will probably
be relevant and we will take them into consideration in future work.

\section{Acknowledgements}

We would like to acknowledge useful comments by N. Furukawa, 
E. Dagotto and G. Aeppli. S.Y. is supported by the
Japanese Society for the Promotion of Science.
A.M. is supported by NSF under grant DMR-95-20776.
Additional support is provided by the National High Magnetic Field Lab 
and MARTECH.
 


\begin{references}
\bibitem{foot} Some authors have attempted to simplify the Hamiltonian by
replacing the spins 3/2 by spins 1 or 1/2 but the quantum effects
associated to small spins are very strong. For this reason our approach
is to replace the spins 3/2 by classical ones which we believe capture
better their physics than smaller spins.
\bibitem{foot3} If antiperiodic BC are used $S(q)$ peaks at $q=0$ in the
ferromagnetic phase. However, in this case $q=\pi/2$, 
which will be relevant later in
this work, is not allowed in the lattice.
\bibitem{foot4} $\Delta \omega=0.1$. The same value was used in all the
figures where spectral data are displayed.
\end{references}
\end{document}